\documentclass[a4paper]{jpconf}
\usepackage{graphicx}

\begin{document}
\title{Quark matter subject to strong magnetic fields: phase diagram
  and applications}

\author{D\'{e}bora P. Menezes, Marcus B. Pinto}
\address{Depto de F\'{\i}sica - CFM - Universidade Federal de Santa
Catarina - Florian\'opolis - SC - CP. 476 - CEP 88.040 - 900 - Brazil}

\author{Constan\c ca Provid\^encia, Pedro Costa, M\'arcio Ferreira}
\address{Centro de F\'{\i}sica Computacional - Department of Physics -
University of Coimbra - P-3004 - 516 - Coimbra - Portugal}

\author{Luis B.  Castro}
\address{Departamento de F\'{\i}sica - Universidade Federal do Maranh\~ao -
S\~ao Lu\'{\i}s - MA - CEP 65080-805 - Brazil}

\ead{debora.p.m@ufsc.br}

\begin{abstract}
In the present work we are interested in understanding various properties of 
quark matter subject to strong magnetic fields described by the 
Nambu--Jona-Lasinio model with Polyakov loop. We start by analysing 
the differences arising from two different vector interactions in the
Lagrangian densities, at zero temperature, and apply the results to 
stellar matter. We then investigate the position of the critical end point for 
different chemical potential and density scenarios.
\end{abstract}

\section{Introduction and Formalism}
\vspace{0.5cm}

The study of the QCD phase diagram, when matter is subject to
strong external magnetic fields has been a topic of intense
investigation recently. The fact that magnetic fields can reach
intensities of the order of $B \sim 10^{19}$ G or higher in
heavy-ion collisions \cite{Skokov:2009qp} and up to $10^{18}$ G in the
center of magnetars \cite{magnetars} made theoretical
physicists consider matter subject to magnetic field both at high
temperatures and low densities and low temperatures and high
densities. 

In this work we study the above mentioned situations within the
framework of the Nambu--Jona-Lasinio model with Polyakov Loop (PNJL) and
different versions of the Nambu--Jona-Lasinio (NJL) model
\cite{njl} for two commonly used parameter sets, known as HK 
\cite{hatsuda} and RKH \cite{reh}.

We describe quark matter subject to strong magnetic fields
within the SU(3) PNJL model with vector interaction and the
Lagrangian density reads:
\begin{eqnarray}
{\cal L} &=& {\bar{\psi}}_f \left[i\gamma_\mu D^{\mu}-
    {\hat m}_f\right ] \psi_f ~+~ {\cal L}_{sym}~+~{\cal L}_{det} \nonumber\\
&+& {\cal L}_{vec}  +{\mathcal{U}} \left(\Phi,\bar\Phi;T\right) - \frac{1}{4}F_{\mu \nu}F^{\mu \nu},
    \label{Pnjl}
\end{eqnarray}
with
\begin{eqnarray*}
    {\cal L}_{sym}&=& G \sum_{a=0}^8 \left [({\bar \psi}_f \lambda_ a \psi_f)^2 +
    ({\bar \psi}_f i\gamma_5 \lambda_a \psi_f)^2 \right ] ,\\
   {\cal L}_{det}&=&-K\left\{{\rm det}_f \left [{\bar \psi}_f(1+\gamma_5)\psi_f \right] +
    {\rm det}_f\left [{\bar \psi}_f(1-\gamma_5)\psi_f\right] \right \},
\end{eqnarray*}
where $\psi_f = (u,d,s)^T$ represents a quark field with three flavors,
${\hat m}_c= {\rm diag}_f (m_u,m_d,m_s)$ is the corresponding (current) 
mass matrix, $\lambda_0=\sqrt{2/3}I$  where $I$ is the unit matrix in the three 
flavor space, and $0<\lambda_a\le 8$ denote the Gell-Mann matrices.
The coupling between the magnetic field $B$ and quarks, and between the
effective gluon field and quarks is implemented  {\it via} the covariant 
derivative $D^{\mu}=\partial^\mu - i q_f A_{EM}^{\mu}-i A^\mu$ where $q_f$ 
represents the quark electric charge, $A^{EM}_\mu=\delta_{\mu 2} x_1 B$ is a 
static and constant magnetic field in the $z$ direction and 
$F_{\mu \nu }=\partial_{\mu }A^{EM}_{\nu }-\partial _{\nu }A^{EM}_{\mu }$.
To describe the pure gauge sector an effective potential
${\mathcal{U}}\left(\Phi,\bar\Phi;T\right)$ is chosen: 
\begin{equation}
 \frac{{\mathcal{U}} \left(\Phi,\bar\Phi;T\right)}{T^4}
  = -\frac{a\left(T\right)}{2}\bar\Phi \Phi \nonumber\\
  +\, b(T)\mbox{ln}\left[1-6\bar\Phi \Phi+4(\bar\Phi^3+ \Phi^3)-3(\bar\Phi \Phi)^2\right],
    \label{Ueff}
\end{equation}
where $a\left(T\right)=a_0+a_1\left(\frac{T_0}{T}\right)+a_2\left(\frac{T_0}{T}\right)^2$, $b(T)=b_3\left(\frac{T_0}{T}\right)^3$.
The standard choice of the parameters for the effective potential $\mathcal{U}$ 
is $a_0 = 3.51$, $a_1 = -2.47$, $a_2 = 15.2$, and $b_3 = -1.75$.

As for the vector interaction, the Lagrangian density that 
denotes the $U(3)_V \otimes U(3)_A$ invariant interaction is 
\begin{equation} 
{\cal L}_{vec} = - G_V \sum_{a=0}^8  
\left[({\bar \psi} \gamma^\mu \lambda_a \psi)^2 + 
 ({\bar \psi} \gamma^\mu \gamma_5 \lambda_a \psi)^2 \right], 
\label{p1} 
\end{equation} 
and  a reduced Lagrangian density can be written as 
 \begin{equation} 
{\cal L}_{vec} = - G_V ({\bar \psi} \gamma^\mu \psi)^2. 
\label{p2} 
\end{equation} 

At zero temperature the PNJL model reduces to the normal NJL model.

In the SU(3) models, the above Lagrangian densities for the vector
sector are not identical  
in a mean field approach and we discuss both cases next. We refer to the 
Lagrangian density given in Eq. (\ref{p1}) as model 1 (P1) and to the
Lagrangian density given in Eq. (\ref{p2}) as model 2 (P2).

For our calculations, 
we need to evaluate the thermodynamical potential for the three flavor 
quark sector, 
$\Omega = -P = {\cal E} - T {\cal S} - \sum_f { \mu_f} \rho_f $ where $P$ 
represents the pressure, ${\cal E}$ the energy density, $T$ the temperature, 
${\cal S}$ the entropy density, and ${\mu_f}$ the chemical potential
of quark with flavor $f$.  
To determine the EOS for the SU(3) NJL at finite density and in the presence 
of a magnetic field in a mean field approximation 
 we need to know the scalar condensates, $\phi_f$, 
the quark number densities, $\rho_f$, as well as 
the pressure kinetic  contribution from the gas of quasi-particles, $\theta_f$.  
In the presence of a magnetic field all these quantities have been evaluated  
with great detail in \cite{prc1,prc2}.

\section{$T=0$ - quark and stellar matter}
\vspace{0.5cm}

\begin{figure}[ht]
\begin{tabular}{cc}
\includegraphics[width=1.\linewidth,angle=0]{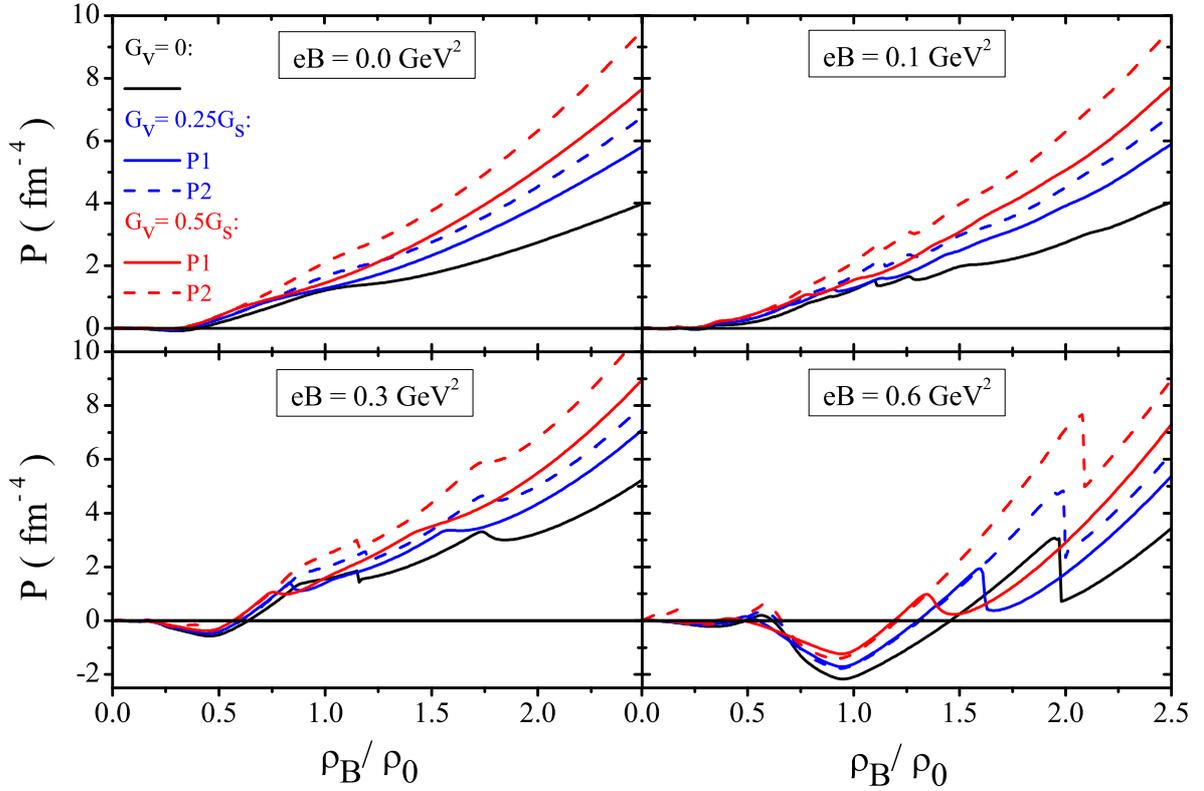}\\
\end{tabular}
\caption{Pressure versus baryonic density for equal chemical
  potentials and models P1 and P2 for different values of $G_V/G_S$,
	and several intensities of the magnetic field. 
	Figure taken from \cite{njl_vetorial}.}
\label{fig1}
\end{figure}

We next restrict ourselves to the NJL model with vector interaction,
disregarding the Polyakov loop, which has shown to be important at
finite temperature only. The effect of the vector interaction on  three flavor 
magnetized matter is studied for cold matter within the two different models
mentioned above, a flavor dependent (P1) \cite{p1} and a flavor
independent one (P2) \cite{p2}. 

If model P1 is considered, the pressure reads: 
\begin{equation}  
P = \theta_u+\theta_d+\theta_s  
-2G_S(\phi_u^2+\phi_d^2+\phi_s^2) +
2G_V ( \rho_u^2 + \rho_d^2 +\rho_s^2) + 4K \phi_u \phi_d \phi_s \,\,, 
\label{pressp1}
\end{equation} 
and the effective chemical potential, for each flavor, is given by  
\begin{equation} 
{\tilde \mu}_i = \mu_i - 4 G_V \rho_i. \quad i=u,d,s 
\label{mup1}
\end{equation} 
We also refer to P1 as the flavor dependent model, for the reasons that
will become obvious from the analysis of our results. 

If, on the other hand,  model P2 is considered, the pressure becomes: 
\begin{equation} 
P = \theta_u+\theta_d+\theta_s  
-2G_S(\phi_u^2+\phi_d^2+\phi_s^2) + 
G_V \rho^2 + 4K \phi_u \phi_d \phi_s \,\,,
\label{pressp2}
\end{equation} 
where 
\begin{equation} 
\rho=\rho_u+\rho_d+\rho_s,  \quad \rho_B=\rho/3, 
\end{equation} 
and in this case the effective chemical potential, for each flavor, 
is given by  
\begin{equation} 
{\tilde \mu}_i = \mu_i - 2 G_V \rho. 
\label{mup2}
\end{equation} 
We next refer to P2 as the flavor independent (or flavor blind)
model.

The effect of the magnetic field on the EOS can be seen in Figure
\ref{fig1} for the case when $\mu_u=\mu_d=\mu_s$.
The van-Alphen oscillations due to the filling of the Landau
levels are already seen for $eB=0.1$ GeV$^2$.  
The softening occurring when a new Landau level
starts being occupied has a strong effect at the smaller densities
giving rise to a pressure that is negative within a larger range of
densities.

\begin{figure}[t]
\begin{tabular}{cc}
\includegraphics[width=0.5\linewidth,angle=0]{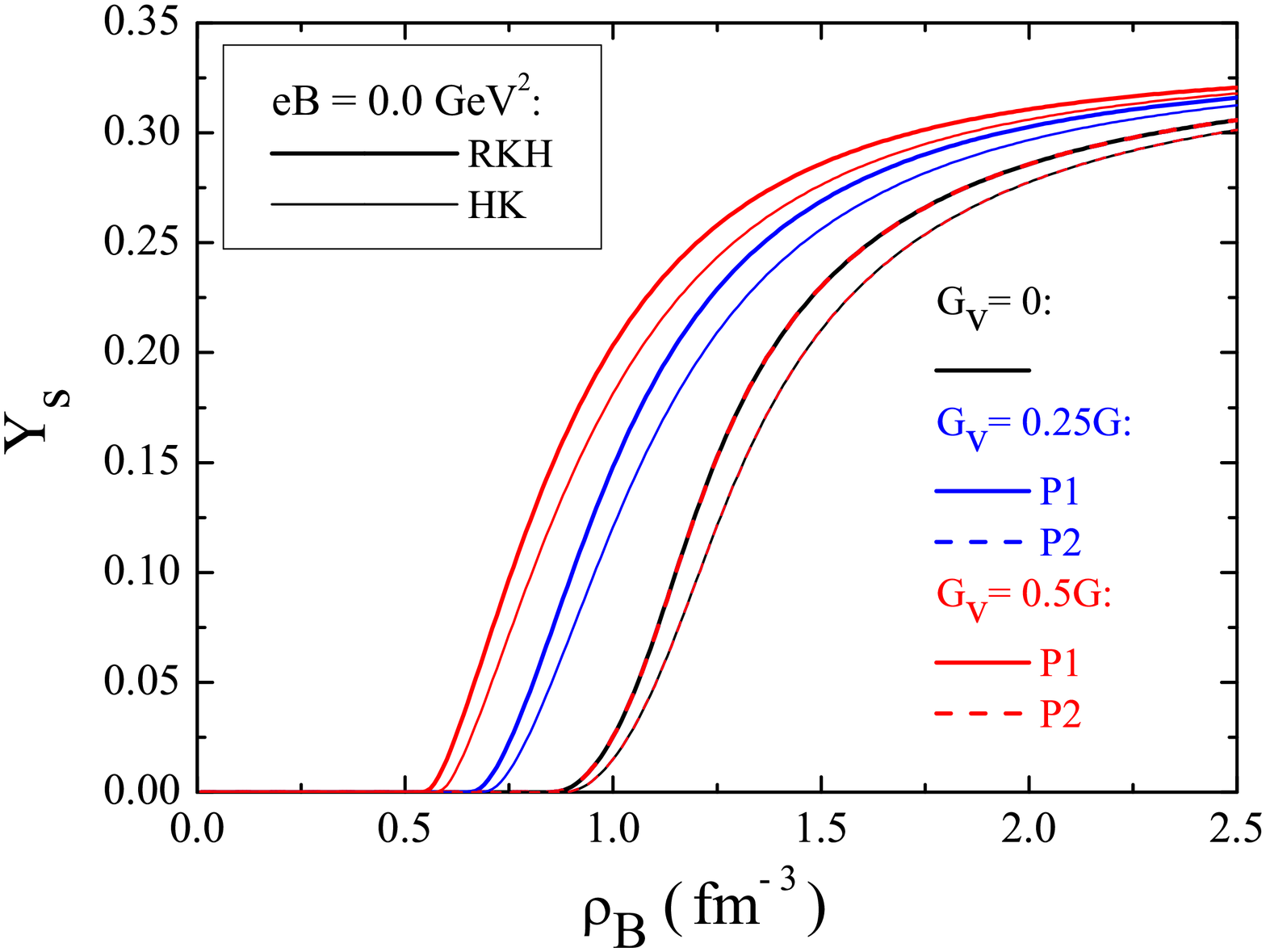} &
\includegraphics[width=0.5\linewidth,angle=0]{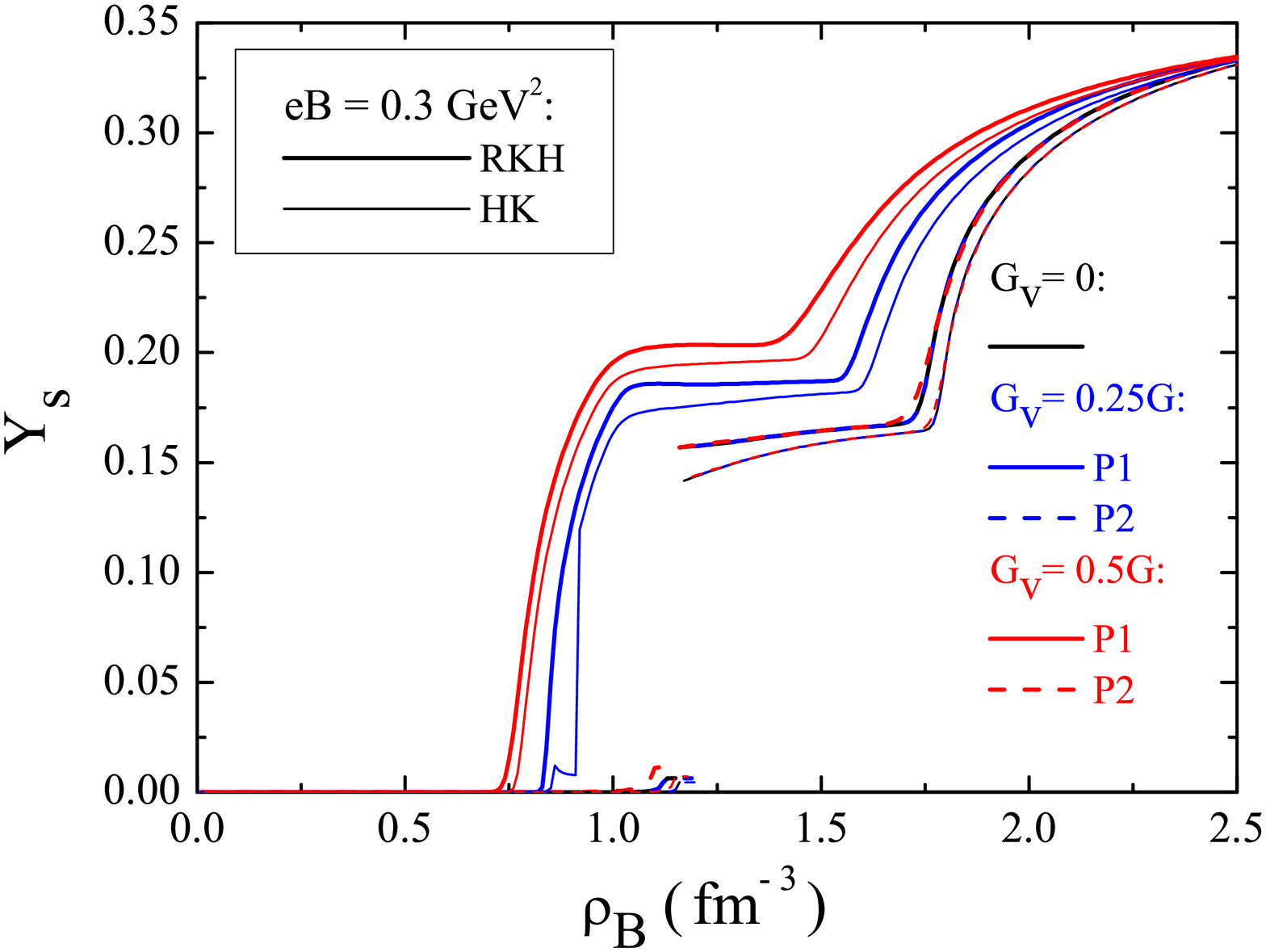} \\
\end{tabular}
\caption{The strangeness fraction as a function of the baryonic
  density  for models P1 and P2 and different values of $G_V$, and $B=0$
	(left panel); $eB=0.3$ GeV$^2$ (right panel).
Figure taken from \cite{njl_vetorial}.}
\label{fig2}
\end{figure}

As can be seen in Figure \ref{fig2}, the
flavor independent vector interaction predicts a smaller strangeness content 
and, therefore, harder equations of state. Moreover, the strangeness
content does not depend on the strength of the vector interaction.
On the other hand, the flavor
dependent vector interaction favors larger strangeness content the
larger the vector coupling,

\begin{figure}[t]
\begin{tabular}{cc}
\includegraphics[width=0.5\linewidth,angle=0]{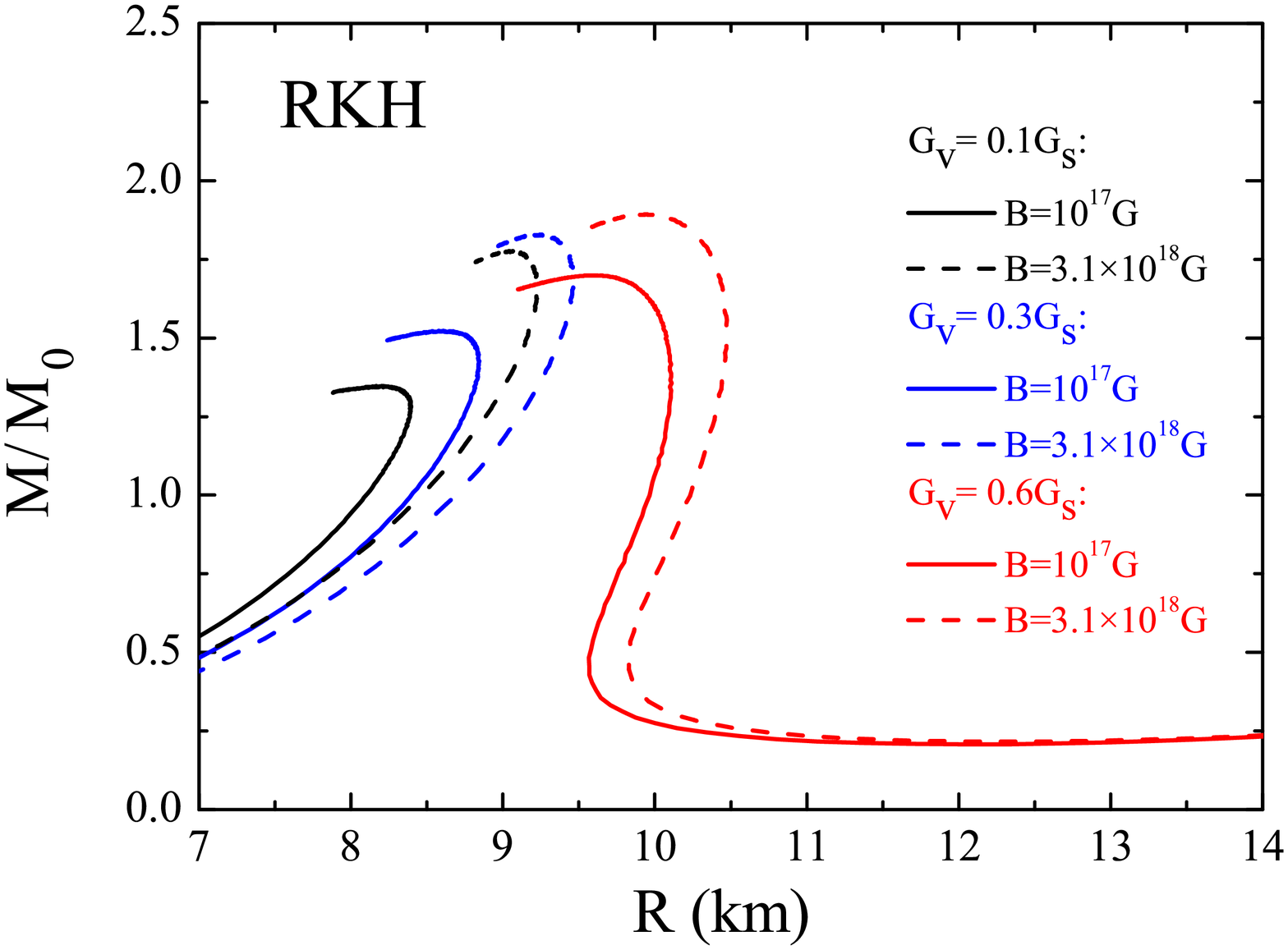} &
\includegraphics[width=0.5\linewidth,angle=0]{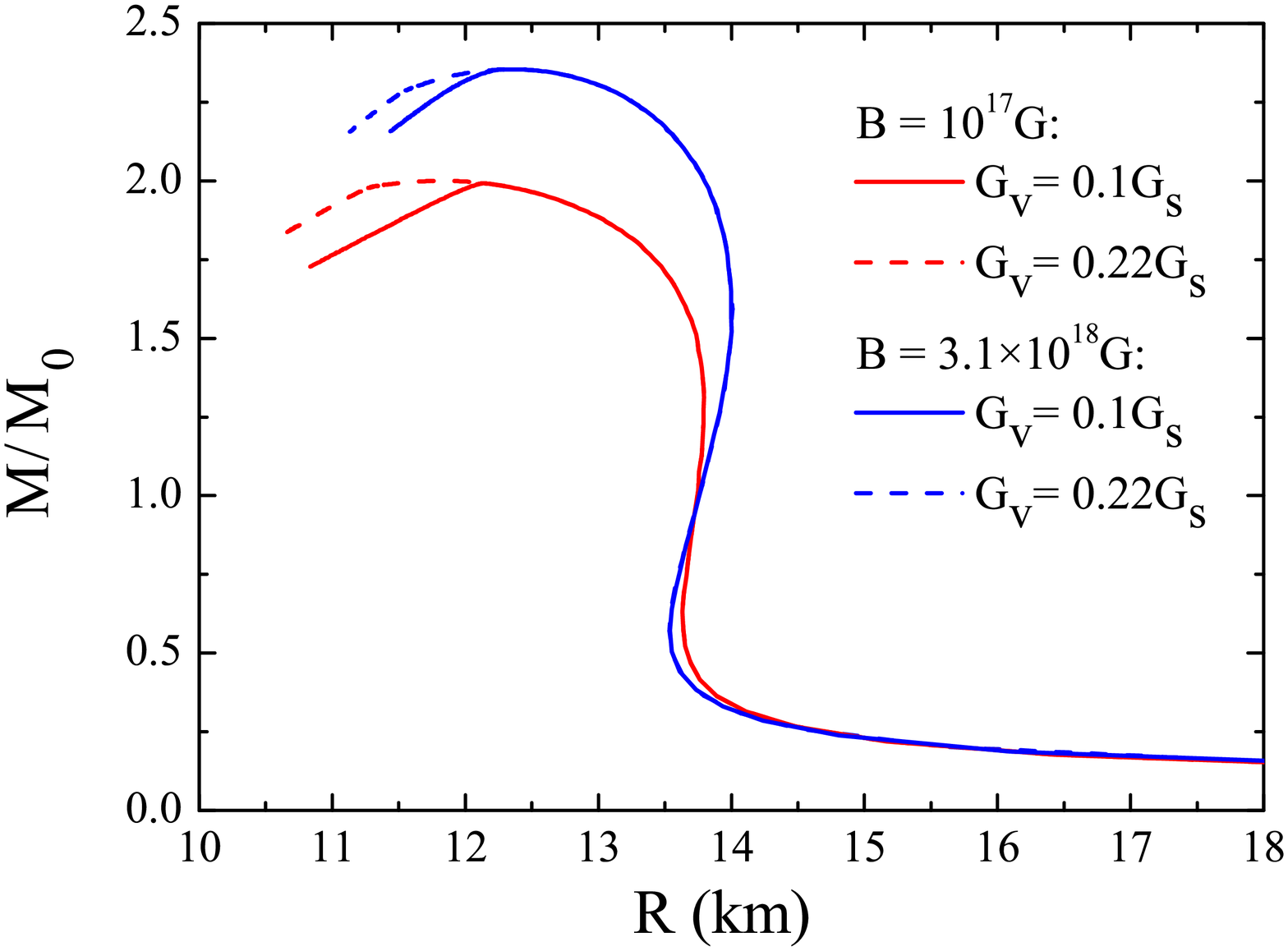} \\
\end{tabular}
\caption{Mass radius curves obtained with P2 for different 
values of $G_V$, two intensities of the magnetic field,
parametrization RKH for quark (left panel) and hybrid stars (right panel).
Figure taken from \cite{njl_vetorial}.}
\label{fig3}
\end{figure}

We now move to the study of quark and hybrid stars subject to an external 
magnetic field. In these cases, $\beta$-equilibrium conditions and charge 
neutrality are enforced.  Mass-radius curves are shown in Figure \ref{fig3} for
both star types with the RKH parameter set. The hadronic phase in
hybrid stars was obtained with the GM1 parametrization of the
non-linear Walecka model \cite{glen}. A Maxwell construction was  
then used.

\begin{table}
\begin{tabular}{|c|c|c|c|c|c|c|c|c|}
\hline
 \multicolumn{3}{ |c| }{} &\multicolumn{3}{ |c| }{HK}&\multicolumn{3}{ |c| }{RKH} \\
  \hline
 \multicolumn{3}{ |c| }{} & $x=0.1$ & $x=0.3$ & $x=0.6$ & $x=0.1$ & $x=0.3$ & $x=0.6$ \\
 \hline
$B=0$~G  & P1 &  $M_{max}$ ($M_{0}$) & 1.49 & 1.58 & 1.69 & 1.27 & 1.35 & 1.46   \\
\cline{3-9}
 &   & R (km) & 9.13 & 10.89 & 11.98 & 8.01 & 8.17 & 9.41 \\
\cline{3-9}
 &   & $\varepsilon_{c}$ ($\mathrm{fm}^{-4}$) & 7.23 & 6.96 & 6.52 & 9.42 & 9.61 & 9.84 \\
\cline{3-9}
\cline{2-9}
 & P2 &  $M_{max}$ ($M_{0}$) & 1.56 & 1.72 & 1.91 & 1.35 & 1.54 & 1.74   \\
\cline{3-9}
 &  & R (km) & 9.15 & 10.61 & 11.47 & 8.22 & 8.60 & 9.91 \\
\cline{3-9}
 &  & $\varepsilon_{c}$ ($\mathrm{fm}^{-4}$) & 7.35 & 7.37 & 6.92 & 8.71 & 8.58 & 8.09 \\
\cline{3-9}
\hline
\hline
\multicolumn{2}{ |c| }{$B=10^{17}$~G} & $M_{max}$ ($M_{0}$) & 1.56 & 1.72 & 1.91 & 1.35 & 1.54 & 1.74 \\
\cline{3-9}
\multicolumn{2}{ |c| }{P2} & R (km) & 9.16 & 10.16 & 10.95 & 8.21 & 8.58 & 9.60 \\
\cline{3-9}
\multicolumn{2}{ |c| }{} & $\varepsilon_{c}$ ($\mathrm{fm}^{-4}$) & 7.41 & 7.36 & 6.98 & 8.80 & 8.94 & 8.11 \\
\cline{3-9}
\hline
\hline
\multicolumn{2}{ |c| }{$B=3.1\times10^{18}$~G} & $M_{max}$ ($M_{0}$) & 1.96 & 2.03 & 2.12 & 1.81 & 1.88 & 1.98 \\
\cline{3-9}
\multicolumn{2}{ |c| }{P2} & R (km) & 9.98 & 10.43 & 11.05 & 9.03 & 9.21 & 9.90 \\
\cline{3-9}
\multicolumn{2}{ |c| }{} & $\varepsilon_{c}$ ($\mathrm{fm}^{-4}$) & 7.41 & 7.22 & 6.78 & 8.74 & 8.21 & 7.80 \\
\cline{3-9}
\hline
\end{tabular}
\caption{Stellar macroscopic properties obtained from EOS of 
non-magnetized matter for models P1 and P2 and for magnetized matter with model
P2. $M_{max}$ is the maximum mass, R is the star radius, $\varepsilon$ the star 
central energy density and $x =G_V/G_S$.
Table taken from \cite{njl_vetorial}.}
\label{table1}
\end{table}

From Table \ref{table1}  we can see that larger star masses
are obtained for the flavor independent vector interaction and maximum
masses of the order of 2 $M_\odot$ can be achieved depending on the value of the 
vector interaction and on the intensity of the magnetic field.

\begin{table}[ht]
\begin{tabular}{|c|c|c|c|c|c|c|c|c|}
 \hline
\multicolumn{2}{ |c| }{HK} & $M_{max}$ & $M_{b}$ & R & $\varepsilon_{c}$ & $\varepsilon$ (onset)  & $\mu_{B}(\varepsilon_{c})$ & $\mu_{B}$ (onset) \\
 \multicolumn{2}{ |c| }{} & ($M_{0}$) & ($M_{0}$) & (km) & ($\mathrm{fm}^{-4}$) & ($\mathrm{fm}^{-4}$)  & (MeV) & (MeV) \\
\hline $B=10^{17}$~G & $x=0$ & 1.91 & 2.18 & 12.78 & 4.57 & 3.47  & 1360 & 1330 \\
\cline{2-9}
P2 & $x=0.10$ & 1.99 & 2.30 & 12.14 & 6.27 & 5.05  & - & 1503 \\
\cline{2-9}
& $x=0.22$ & 2.00 & 2.31 & 11.82 & 5.93 & 7.79  & 1580 & 1726 \\
\hline
 $B=3.1\times10^{18}$~G & $x=0$ & 2.27 & 2.60 & 12.82 & 4.69 & 3.30  & 1324 & 1261 \\
\cline{2-9}
P2 & $x=0.10$ & 2.35 & 2.70 & 12.34 & 5.29 & 5.59 & 1427 & 1453 \\
\cline{2-9}
& $x=0.22$ & 2.35 & 2.70 & 12.35 & 5.27 & 9.03 & 1426 & 1730 \\
\hline
 \hline
 \multicolumn{2}{ |c| }{RKH} & $M_{max}$ & $M_{b}$ & R &
 $\varepsilon_{c}$ & $\varepsilon$ (onset)  & 
$\mu_{B}(\varepsilon_{c})$ & $\mu_{B}$ (onset) \\
 \multicolumn{2}{ |c| }{} & ($M_{0}$) & ($M_{0}$) & (km) & ($\mathrm{fm}^{-4}$) & ($\mathrm{fm}^{-4}$)  & (MeV) & (MeV) \\
\hline $B=10^{17}$~G & $x=0$ & 1.97 & 2.26 & 12.48 & 4.29 & 4.28 & - & 1422 \\
\cline{2-9}
P2 & $x=0.10$ & 2.00 & 2.31 & 11.91 & 7.51 & 5.67 & - & 1557 \\
\cline{2-9}
& $x=0.19$ & 2.00 & 2.31 & 11.83 & 5.91 & 7.83  & 1579 & 1728 \\
\hline
 $B=3.1\times10^{18}$~G & $x=0$ & 2.33 & 2.69 & 12.79 & 4.69 & 4.19 &  - & 1335 \\
\cline{2-9}
P2 & $x=0.10$ & 2.35 & 2.70 & 12.34 & 5.30 & 6.52 & 1428 & 1531 \\
\cline{2-9}
& $x=0.19$ & 2.35 & 2.70 & 12.34 & 5.30 & 9.05 & 1428 & 1731 \\
\hline
\end{tabular}
\caption{Stellar macroscopic properties obtained from EOS of
  magnetized hybrid stars built with GM1 and SU(3) NJL with HK and RKH
  parametrizations. $M_{max}$ is the maximum gravitational mass,
  $M_{b}$ is the maximum baryonic mass, $R$ is the star radius,
  $\varepsilon_{c}$ is the star central energy density,
  $\mu_{B}(\varepsilon_{c})$ is the chemical potential for neutron at
  $\varepsilon_{c}$ and $\mu_{B}$(onset) is the baryonic chemical
  potential at the onset of the quark phase. 
Results taken from \cite{njl_vetorial}.}
\label{tableh}
\end{table}

From the results displayed in Table \ref{tableh}, we observe that
hybrid stars may bare a core containing deconfined quarks if neither the vector
interaction nor the magnetic field are  too strong. The onset of the
quark phase as compared with the central values for the energy density
and related chemical potential show that quarks would appear only at
densities/chemical potentials larger than the existing ones in the
stellar core.
The presence of strong magnetic fields also disfavor the existence of a quark 
core in hybrid stars. 

\section{Critical end point within the PNJL model} 
\vspace{0.5cm}

\begin{figure}[t]
\begin{tabular}{cc}
\includegraphics[width=0.5\linewidth,angle=0]{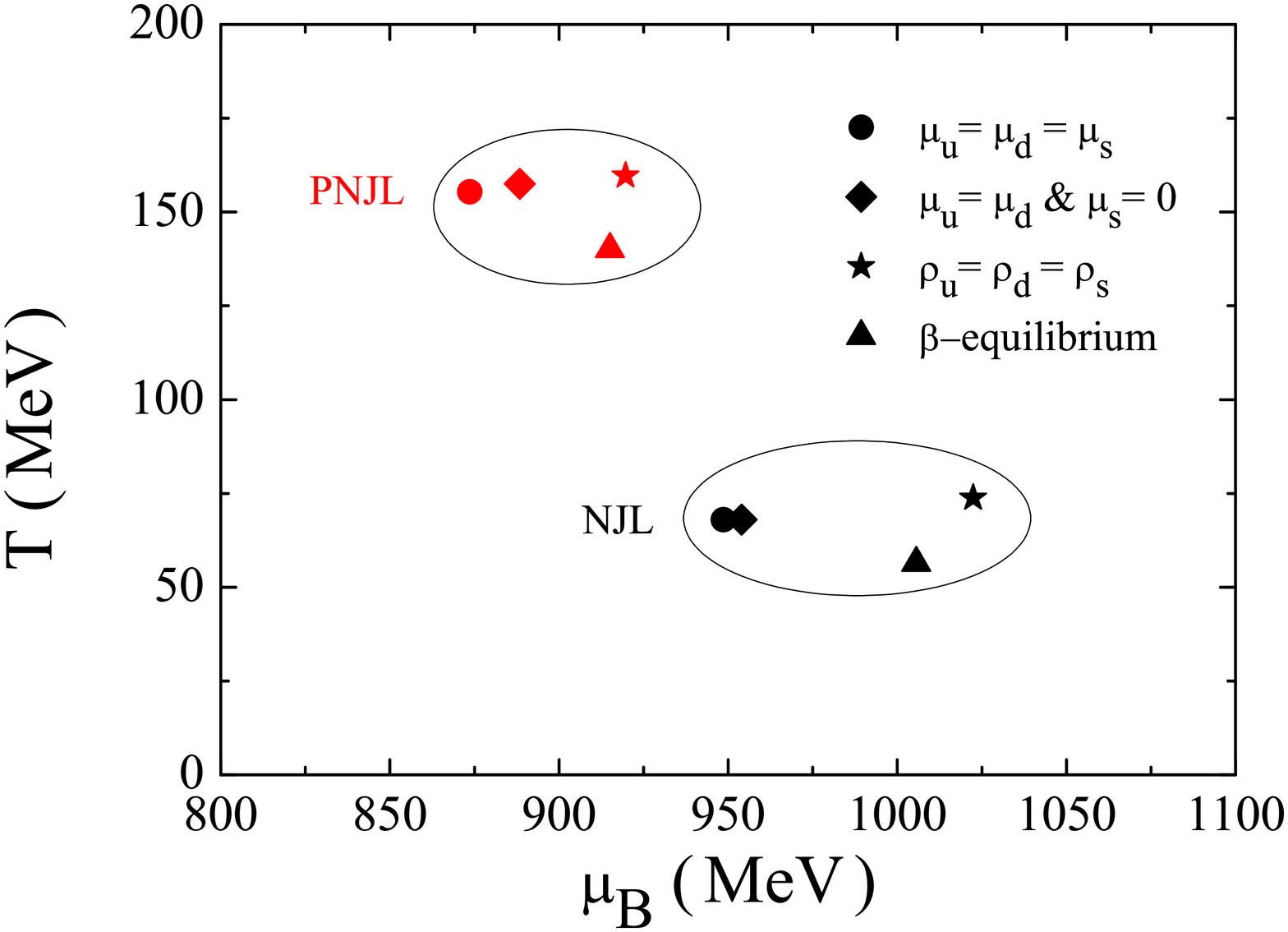} &
\includegraphics[width=0.5\linewidth,angle=0]{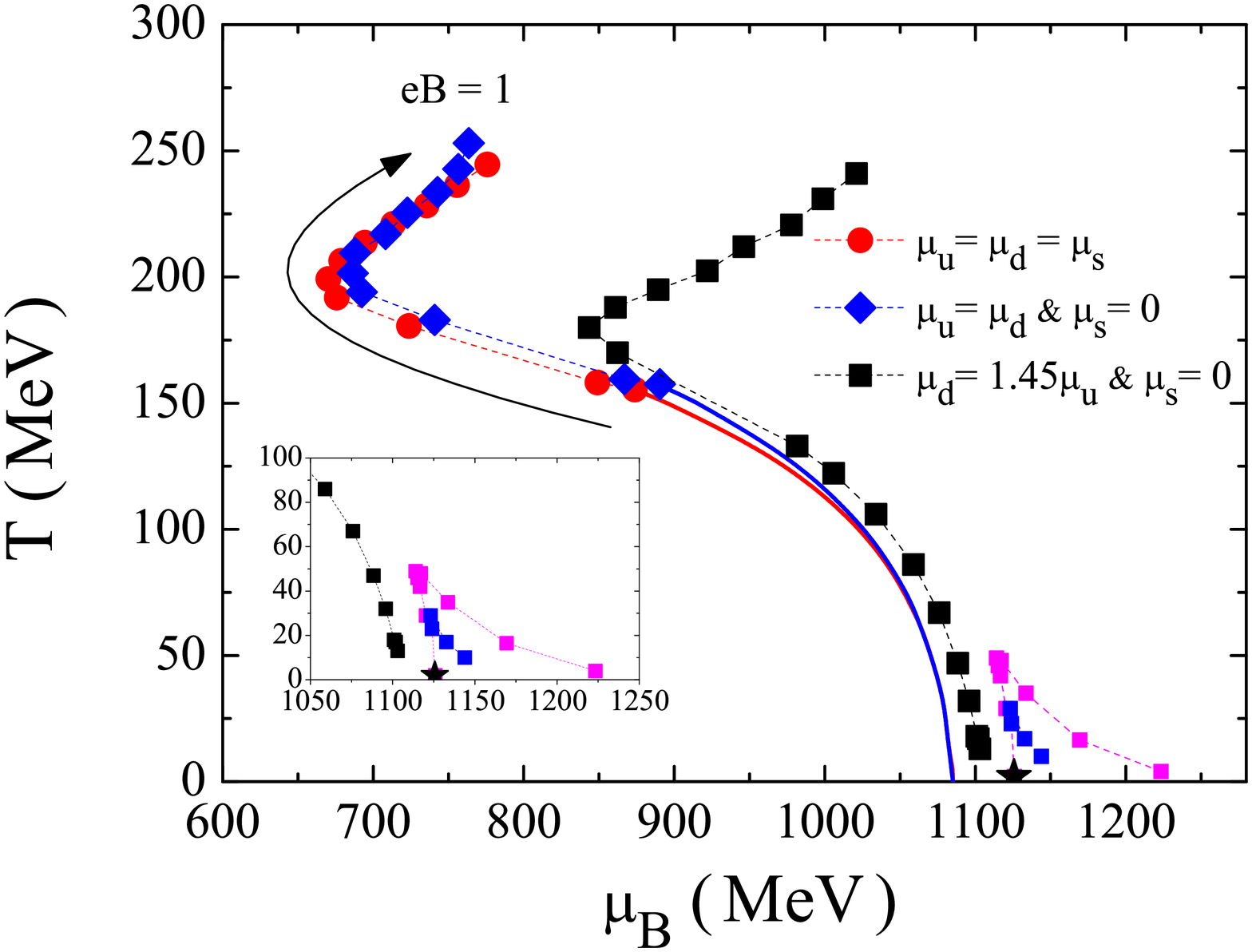}\\
\end{tabular}
\caption{Left panel - Location of the CEP on a diagram $T$ vs the baryonic
chemical potential under different scenarios and
models (NJL, PNJL). No external magnetic field is considered.
Right panel - Effect of an external magnetic field on the CEP location
within PNJL model for three different scenarios.
Figure taken from \cite{cep}.} 
	\label{fig5}
\end{figure}

We move next to finite temperature and study the location of the critical end 
point (CEP) on the QCD phase diagram within the framework of the PNJL model and 
without the introduction of the vector interaction.
We investigate different scenarios with respect to the isospin and strangeness 
content of matter, as shown in Figure \ref{fig5} left for non-magnetized matter.  
For matter in $\beta$-equilibrium, the CEP occurs at smaller temperatures and
densities. This scenario is of interest for neutron stars and confirms previous 
calculations that indicate that a deconfinement phase transition in the 
laboratory will be more easily attained with asymmetric nuclear matter
\cite{Cavagnoli:2010yb}. 

As a matter of fact, by taking $\mu_s=0$ and increasing systematically 
$\mu_d$ with respect to $\mu_u$ the CEP moves to smaller temperatures and 
larger baryonic chemical potentials \cite{cep}: 
less symmetric matter is less bound and, therefore, the transition to a 
chirally symmetric phase occurs at a smaller temperature and density than in 
the symmetric case ($\mu_u=\mu_d=\mu_s$). 
When the asymmetry is large enough, $\mu_d = 1.45 ~\mu_u$, the CEP disappears. 

Another interesting situation is observed when analyzing very isospin 
asymmetric matter subject to different intensities of
the magnetic field, as seen in Figure \ref{fig5} right.
Starting from a scenario having an isospin asymmetry above which the
CEP does not exist for a zero external magnetic field ($\mu_d = 1.45~\mu_u$) 
it was shown the introduction of an
external magnetic field could drive the system to a first order phase 
transition. The CEP occurs at very small temperatures if $eB<0.1$ GeV$^2$ and, 
in this case, a complicated structure with several CEP at different values of
($T, \mu_B)$ are possible for the same magnetic field, because the
temperature is not high enough to wash out the Landau level effects. 
Indeed, the magnetic field affects in a different way $u$ and $d$ quarks due to 
their different electric charge. A consequence is the possible appearance of 
two or more CEPs for a given magnetic field intensity.
For $eB>0.1$ GeV$^2$ only one CEP exists.
This is an important result because it shows that an external magnetic
field is able to drive a system with no CEP into a first order phase
transition \cite{cep}.

In the right panel of Figure \ref{fig5} we present the results for two 
other scenarios where a first order phase transition already occurs at zero 
magnetic field: $\mu_u=\mu_d=\mu_s$ and $\mu_u=\mu_d\,\&\,\mu_s=0$.
For moderate magnetic fields the trend is very similar for all scenarios: 
as the intensity of the magnetic field increases, the transition temperature 
increases and the baryonic chemical potential decreases until the 
critical value $eB \sim 0.4$ GeV$^2$.
For stronger magnetic fields both $T$ and $\mu_B$ increase.

It is also important to point out that all three scenarios presented in 
the right panel of Figure \ref{fig5} show an inverse magnetic catalysis at 
finite chemical potential and zero temperature, i.e., the critical temperature 
decreases with increasing $eB$. 
For large values of $eB$ this inverse magnetic catalysis tendency disappears 
and a magnetic catalysis takes place.

\vspace{0.5cm}
{\bf Acknowledgements:} This work was partially supported by CNPq (Brazil), 
CAPES (Brazil) and FAPESC (Brazil) under Project 2716/2012,TR 2012000344,
by Project No. PTDC/FIS/113292/2009 developed under the initiative QREN 
financed by the UE/FEDER through the program COMPETE —''Programa
Operacional Factores de Competitividade,'' by Grant No. SFRH/BD/51717/2011
(F.C.T.) and by NEW COMPSTAR, a COST initiative.

\end{document}